\documentstyle[12pt,epsfig]{article}
\advance \topmargin by -\headheight
\advance \topmargin by -\headsep
\evensidemargin \oddsidemargin
\begin{document}
\pagestyle{plain}
\begin{titlepage}
\vspace*{0.15cm}
\begin{center}
{\large\bf INSTITUTE FOR HIGH ENERGY PHYSICS}
\end{center}
\vspace*{1.cm}


\vspace*{3.cm}

\begin{center}
{\Large\bf
   High statistic study of the $K^{-} \rightarrow \pi^{0} \mu^{-} \nu $ 
    decay} \\

\vspace*{0.15cm}
\vspace*{1.3cm}

{\bf   O.P.~Yushchenko, S.A.~Akimenko,  K.S.~Belous, 
 G.I.~Britvich, I.G.Britvich, K.V.Datsko,  A.P.~Filin, 
A.V.~Inyakin,  A.S.~Konstantinov, V.F.~Konstantinov,  
I.Y.~Korolkov, V.A.~Khmelnikov, V.M.~Leontiev, V.P.~Novikov,
V.F.~Obraztsov,  V.A.~Polyakov, V.I.~Romanovsky, V.M.~Ronjin, 
   V.I.~Shelikhov, N.E.~Smirnov,
  O.G.~Tchikilev, V.A.Uvarov. }
\vskip 0.15cm
{\large\bf $Institute~for~High~Energy~Physics,~Protvino,~Russia$}
\vskip 0.35cm
{\bf V.N.~Bolotov, S.V.~Laptev, A.R.~Pastsjak, A.Yu.~Polyarush. }
\vskip 0.15cm
{\large\bf $Institute~for~Nuclear~Research,~Moscow,~Russia$}
\end{center}

\vspace*{3cm}

\begin{abstract}
 The  decay $K^{-} \rightarrow \pi^{0} \mu^- \nu$ has been 
 studied using in-flight decays detected with the "ISTRA+" spectrometer. 
 About 540K events were
 collected for the analysis.  
 The $\lambda_{+}$  and  $\lambda_{0}$ slope
 parameters of the decay form-factors  $f_{+}(t)$, $f_{0}(t)$
  have been measured : 
 $\lambda_{+}= 0.0277 \pm 0.0013$(stat) $\pm 0.0009$(syst),  
 $\lambda_{0}= 0.0183 \pm 0.0011$(stat) $\pm 0.0006$(syst), and 
 $d \lambda_{0}/d \lambda_{+}=-0.348$.
  The limits on the
 possible tensor and scalar couplings have been derived:
 $f_{T}/f_{+}(0)=-0.0007 \pm 0.0071$, 
 $f_{S}/f_{+}(0)=0.0017 \pm 0.0014$.
 No visible non-linearity in the form-factors have been observed.
 \end{abstract}

\end{titlepage}

\newpage
\thispagestyle{empty}

~

\setcounter{page}{0}
\newpage
\raggedbottom
\sloppy

\section{ Introduction}

 The decay  $K \rightarrow \mu \nu \pi^{0}$(K$_{\mu 3}$) provides unique
 information about the dynamics of the strong interactions. It has been a
 testing ground for such theories as current algebra, PCAC, Chiral Perturbation
 Theory(ChPT). In this paper we present a high-statistics measurement
 ($\sim$~537K events) of the Dalitz plot density in this decay. 
 This study has a
 particular interest in view of new two-loop order ($p^6$) calculations for
 K$_{l 3}$ in ChPT \cite{Bijnens}.
 
 The K$_{\mu 3}$ decay is also known to be a key 
  one in hunting for phenomena beyond the Standard Model (SM).
 In  particular, significant efforts have been invested into T-violation searches,
 by the measurements of the muon transverse polarization $\sigma_{T}$ 
 \cite{KEK1}, as well as into searches for the non-SM contributions into 
 the decay amplitude \cite{KEK2}. 

 In our analysis  
 we present new search for 
 scalar (S) and tensor (T) interactions  by fitting the $K_{\mu 3}$ 
 Dalitz plot
 distribution, similar to the procedure used in the $K_{e3}$ decay 
 studies\cite {papere}.
   
  
\section{ Experimental setup}
The experiment has been performed at the IHEP 70 GeV proton synchrotron U-70.
The experimental setup "ISTRA+" (Fig.1) 
was described in some details in our paper  \cite{Paper1}. 

\begin{figure}[h]
\epsfig{file=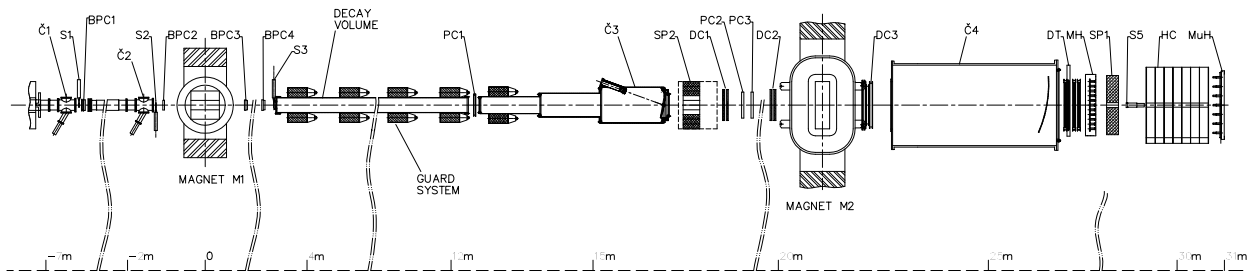, angle=90 ,width=16.5cm }
\caption{ Elevation view of the   "ISTRA+" detector.}
\end{figure}


 The setup is located in a negative unseparated secondary beam. 
The beam momentum is $\sim 25$ GeV with 
$\Delta p/p \sim 1.5 \% $. The admixture of $K^{-}$ in the beam is $\sim 3 \%$.
The beam intensity is $\sim 3 \cdot 10^{6}$ per 1.9 sec. of the U-70 spill.
The  beam particles are deflected by the beam magnet M$_{1}$ and are 
measured by $BPC_{1}\div BPC_{4}$ proportional chambers
with 1~mm wire spacing. The kaon identification is performed by 
$\check{C_{0}} \div \check{C_{2}}$ threshold $\check{C}$-counters. 

The 9 meter long vacuumed
decay volume is surrounded by 8 lead-glass rings $LG_{1} \div LG_{8}$ 
which are used as the veto system for 
low energy photons. The photons radiated at large angles are detected 
by the lead-glass calorimeter $SP_{2}$.

The decay products are deflected by the spectrometer magnet  M2 with 
a field integral of 1~Tm. The track measurement is performed by 
 2-mm-step proportional chambers  
($PC_{1} \div PC_{3}$), 1-cm-cell drift chambers  
($DC_{1} \div DC_{3}$),
 and by 2-cm-diameter
drift  tubes    ($DT_{1} \div DT_{4}$). 
Wide aperture threshold \v{C}erenkov counters ($\check{C_{3}}$ and 
$\check{C_{4}}$) are filled  with helium and
are not used in these measurements. 

The photons are measured by the lead-glass calorimeter $SP_{1}$ which consists
of 576 counters. The counter transverse  
size is $5.2\times 5.2$ cm and the 
length is about 15~$X_0$. 

The  scintillator-iron sampling hadron calorimeter HC is subdivided
into 7 longitudinal sections 7$\times$7 cells each. The 11$\times$11 cell 
scintillating hodoscope is used for the improvement of the time
resolution of the tracking system. 
MuH  is a 7$\times$7 cell scintillating muon hodoscope.

The trigger is provided by $S_{1} \div S_{5}$ scintillation counters, 
$\check{C_{0}} \div \check{C_{2}}$ Cerenkov counters, and
the analog sum of amplitudes from last dinodes of the $SP_1$ :
\begin{displaymath}
 T=S_{1} \cdot S_{2} \cdot S_{3} \cdot 
 \bar{S_{4}} \cdot \check{C_{0}} \cdot \bar{\check{C_{1}}} \cdot 
 \bar{\check{C_{2}}} \cdot 
 \bar{S_{5}} \cdot \Sigma(SP_{1}),
\end{displaymath}
where $S_4$ is the scintillator counter with a hole to suppress the beam halo,
 $S_5$ is the counter located downstream  the setup at the beam focus. This part
 of the trigger is intended to identify beam kaons and to kill undecayed
 particles. It is designed on purpose, in a very simple way, to avoid any bias.
$\Sigma(SP_{1})$ requires that the analog sum of amplitudes from 
the $SP_1$ be larger than $\sim$700 MeV - the MIP signal. The last requirement 
serves to suppress the dominating $K \rightarrow \mu \nu$ decay. A part of
events (10\%) which do not satisfy the 
$\Sigma(SP_{1})$ requirement is also recorded
to provide the information for muon identification studies.

\section{Events selection}

During run in Winter 2001, 
332M events were logged on tapes.
This statistics is complemented by about 130M MC events generated with 
Geant3 \cite{geant} Monte Carlo program. The MC generation
includes a realistic description of the setup with decay volume 
entrance windows,
tracking chambers windows, chambers gas mixtures, sense wires and cathode 
structures,
\v{C}erenkov counters mirrors and gas, the shower generation in EM calorimeters, 
etc.

The data processing starts with the beam particle reconstruction in 
$BPC_{1} \div BPC_{4}$. Then secondary tracks are looked for in the
decay tracking system
and events with one good negatively charged track are selected.
The decay vertex is reconstructed by means of the 
unconstrained vertex fit of the beam and decay tracks.

A clustering procedure is used to find showers in the 
$SP_{1}$ calorimeter, and the two-dimensional pattern of the shower is fitted 
with the MC-generated patterns to 
reconstruct its energy and position.

The muon identification is done using the information from electromagnetic and
hadronic calorimeters. First of all, the energy deposition in the 
$SP_1$ associated
with the track (counted in the 3x3 matrix around the track extrapolation to
the $SP_1$) is required to be less than 500 MeV. This cut is intended to
suppress the electron tracks. 
The sum of ADC counts 
from the HC counters associated with 
remaining tracks is demanded to be less than 200 (see Figure 2). 
And, finally, the ratio of the associated ADC signals in the last
three layers of HC to the total associated ADC sum to be greater
than 0.05 is required 
(Figure 3 shows this value for the tracks which pass the first two
selection criteria).

\begin{minipage}[t]{8.cm}
\epsfig{file=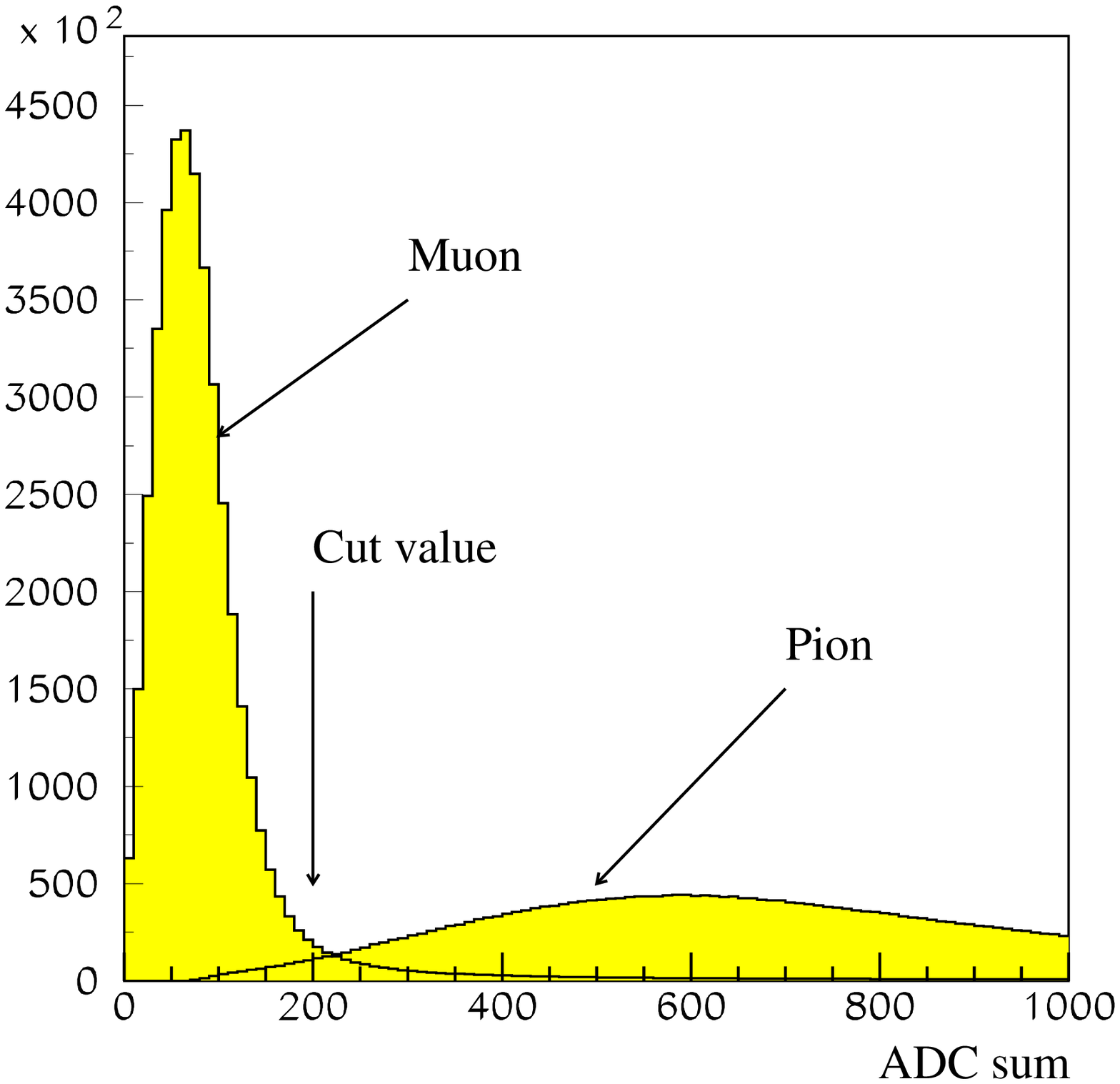,width=8cm}

\begin{center}
Figure 2: The ADC sum in HC for the track-associated cells.
\end{center}
\end{minipage} \ \hfill \ 
\begin{minipage}[t]{8.cm}
\epsfig{file=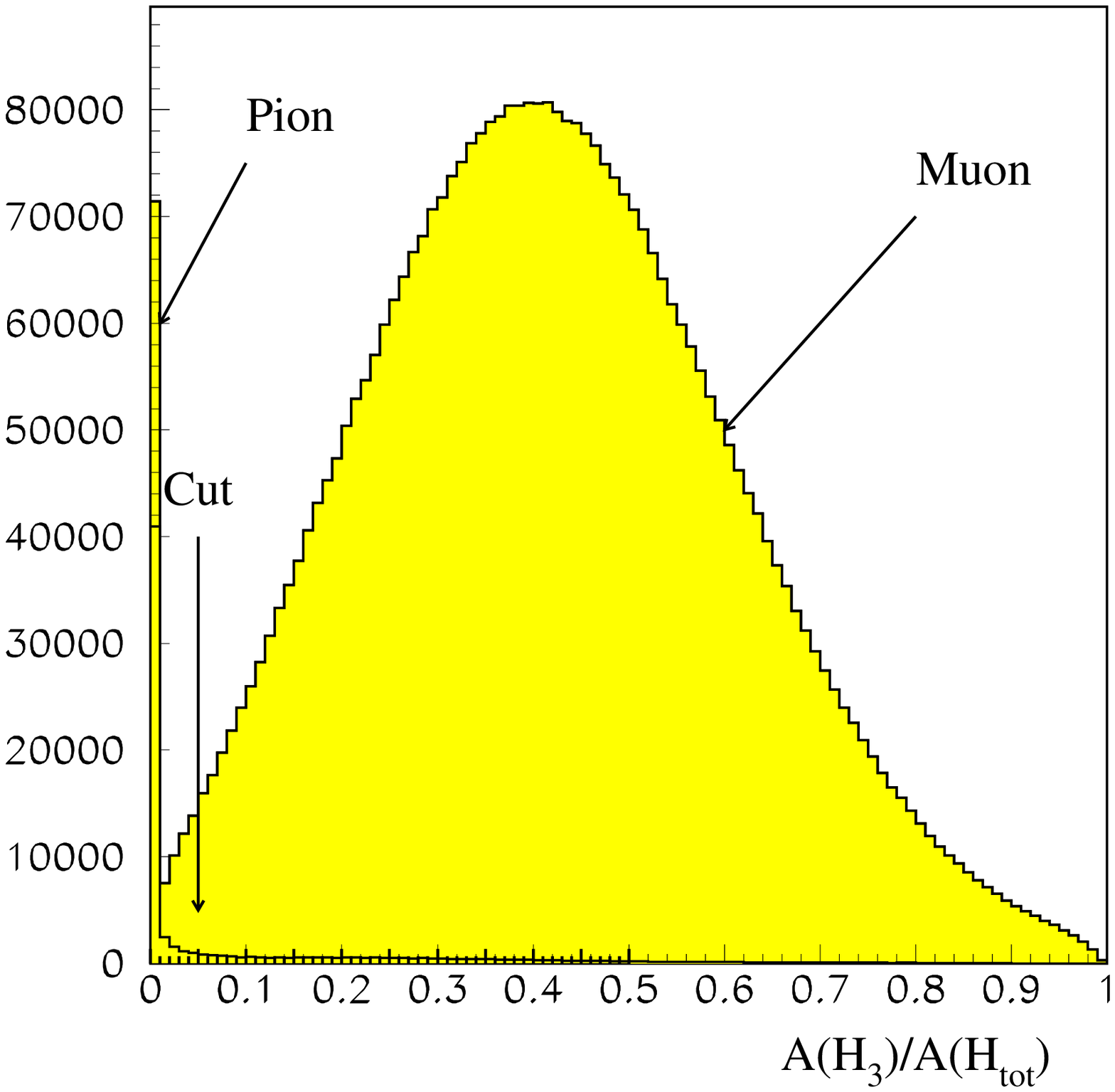,width=8cm}
\begin{center}
Figure 3: The ratio of the track-associated ADC signals in the last three 
layers of HC to the total associated signal. 
\end{center}
\end{minipage}

~

~

 The  figures 2 and 3 are obtained with the clean $\pi^-$ and $\mu^-$ samples 
selected from the real data. The pion data sample is composed from
selected $K^-\to \pi^- \pi^0$ decays, and the muon one from the 
$K^-\to \mu^- \nu_{\mu}$ decays.

  The efficiency of the muon identification and the probability of 
  the $\pi\to\mu$ assignment were found to be 88\%\ and 0.03 respectively.
 
 The events with one charged track identified as muon and two additional 
showers in the $SP_1$ are selected for further processing.

 The selected events are required to pass 2C
 $K \rightarrow \mu \nu \pi^{0}$ fit, with a probability of the fit 
 $P_{\mbox{fit}} > 0.005$.
  The angle between $\pi^0$ and $\mu^-$ 
 in the kaon rest frame after 2C fit was found to be a good variable for
 the further background suppression (see Figure 4). 
 The background from the surviving 
 $K^- \to \pi^- \pi^0$ events is concentrated at $\cos\theta \sim -1$, and 
the selected cut $\cos\theta_{\pi\mu} > -0.95$ removes practically all the
background.
  The missing energy 
 $E_{\nu}=E_{K}-E_{\mu}-E_{\pi^{0}}$ after the angular cut is shown in Figure 5. 
The signal Monte-Carlo events for Figures 4 and 5 are weighted with the 
$K_{\mu 3}$ matrix element where we use  $\lambda_{+} = 0.0286$ (fixed from 
our $K_{e3}$ measurements \cite{papere}) and 
$\lambda_0 = 0.017$ (from the ChPT $\mbox{O}(p^4)$ calculations \cite{Leutwyler}).

 We estimate the 
 surviving background contribution to be around 0.3\%.

\begin{minipage}[t]{8.cm}
\epsfig{file=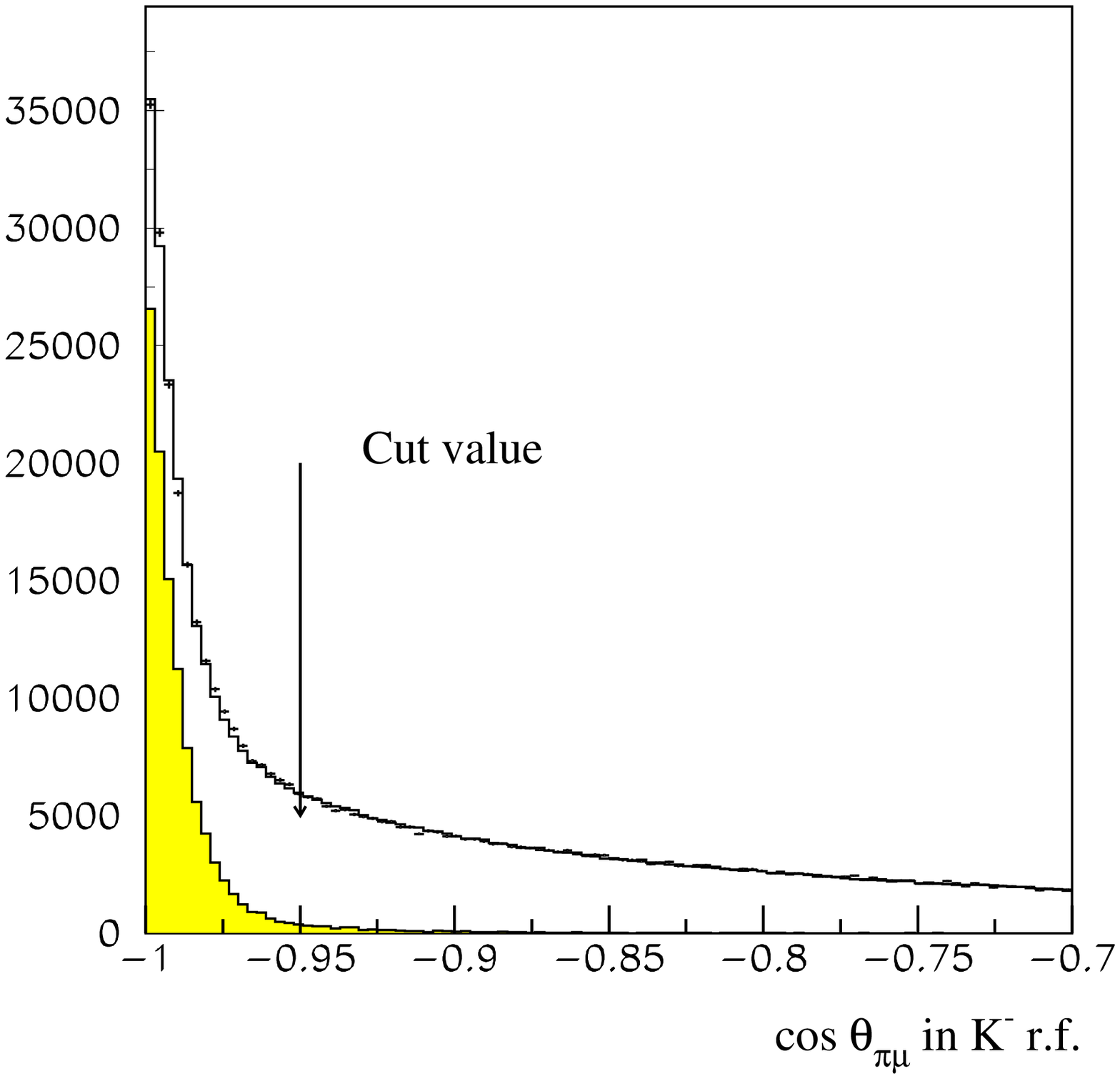,width=8cm}

\begin{center}
Figure 4: The cosine of the $\pi - \mu$ angle in the kaon rest frame after 
2C fit. 
 The points with errors are data and the solid histogram is MC. The shaded 
 area shows the background contribution. 
\end{center}
\end{minipage} \ \hfill \ 
\begin{minipage}[t]{8.cm}
\epsfig{file=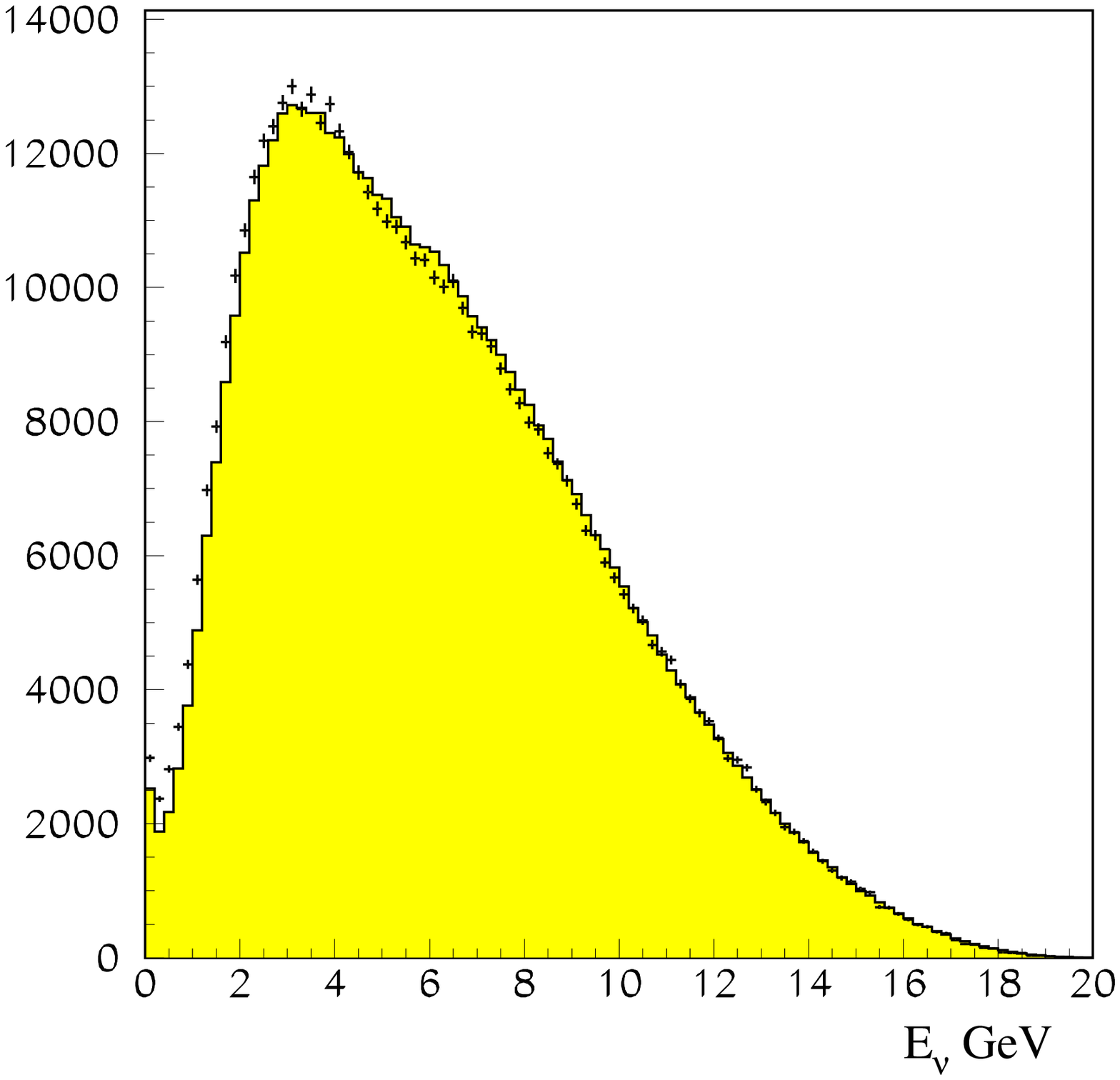,width=8cm}
\begin{center}
Figure 5: The $E_\nu$ compared with MC.
\end{center}
\end{minipage}

~

~
 
\section{ Analysis}

After the procedure described in the previous section,
537K events are selected in the real data. The distribution of the events 
over the Dalitz plot is shown in Figure 6.

The most general Lorentz-invariant form of the matrix element for the 
$K^{-} \rightarrow l^{-} \nu \pi^{0}$ decay is \cite{Steiner}:
\begin{equation}
M= \frac{G_{F}V_{us}}{2} \bar u(p_{\nu}) (1+ \gamma^{5})
[2m_{K}f_{S} -
[(P_{K}+P_{\pi})_{\alpha}f_{+}+
(P_{K}-P_{\pi})_{\alpha}f_{-}]\gamma^{\alpha} + i \frac{2f_{T}}{m_{K}}
\sigma_{\alpha \beta}P^{\alpha}_{K}P^{\beta}_{\pi}]v(p_{l})
\end{equation}
It consists of scalar, vector, and tensor terms. The $f_{\pm}$ form-factors
are the functions of $t= (P_{K}-P_{\pi})^{2}$. In the Standard Model (SM),
the W-boson exchange leads to the pure vector term. 
The scalar and/or tensor terms which are ``induced'' by
 EW radiative corrections are negligibly
small, i.e  nonzero scalar or tensor form-factors would indicate 
the physics beyond the SM. 

The term in the vector part, proportional to $f_{-}$, 
is reduced (using the Dirac
equation) to the scalar form-factor. In the same way, the tensor term is 
reduced to
a mixture of the scalar and  vector form-factors. The redefined vector (V) and 
scalar (S) terms, and the corresponding Dalitz plot
density in the kaon rest frame ($\rho(E_{\pi},E_{l})$) are \cite{Chizov}:
\begin{eqnarray}
\rho (E_{\pi},E_{l}) & \sim & A \cdot |V|^{2}+B \cdot Re(V^{*}S)+C \cdot |S|^{2} \\
 V & = & f_{+}+(m_{l}/m_{K})f_{T} \nonumber \\ 
S & = & f_{S} +(m_{l}/2m_{K})f_{-}+
\left( 1+\frac{m_{l}^{2}}{2m_{K}^{2}}-\frac{2E_{l}}{m_{K}}
-\frac{E_{\pi}}{m_{K}}\right) f_{T} \nonumber \\ 
A & = & m_{K}(2E_{l}E_{\nu}-m_{K} \Delta E_{\pi})-  
m_{l}^{2}(E_{\nu}-\frac{1}{4} \Delta E_{\pi}) \nonumber \\
B & = & m_{\l}m_{K}(2E_{\nu}-\Delta E_{\pi}) ;~ E_{\nu}=m_{K}- E_{l}-E_{\pi}
 \nonumber \\
C & = & m_{K}^{2} \Delta E_{\pi};~ \Delta E_{\pi}  =  E_{\pi}^{max}-E_{\pi} ;~
E_{\pi}^{max}= \frac{m_{K}^{2}-m_{l}^{2}+m_{\pi}^{2}}{2m_{K}} \nonumber 
\end{eqnarray}

Following \cite{Leutwyler} the scalar form-factor $f_{0}$ is introduced:
\begin{equation}
f_{0}(t)=f_{+}(t)+ \frac{t}{m_{K}^{2}-m_{\pi}^{2}}f_{-}(t),
\end{equation}
 and we assume, at most, 
the quadratic dependence of $f_{+},\; f_{0}$ on t:
\begin{equation}
 f_{+}(t)=f_{+}(0)\left( 1+\lambda_{+} t/m_{\pi}^2+\lambda_{+}^{'}t^2/m_{\pi}^4 
 \right),\;\;\;
 f_{0}(t)=f_{+}(0)\left( 1+\lambda_{0} t/m_{\pi}^2+\lambda_{0}^{'}t^2/m_{\pi}^4 
 \right).
\end{equation}
 Finally, one gets from Eq. (3):
\begin{equation} 
 f_{-}=f_{+}(0)\frac{m_{K}^{2}-m_{\pi}^{2}}{m_{\pi}^{2}}\cdot
 \left( \lambda_0 - \lambda_{+} + \frac{t}{m_{\pi}^4} (\lambda_{0}^{'} - 
 \lambda_{+}^{'})\right)
\end{equation}
  
\begin{center}
\epsfig{file=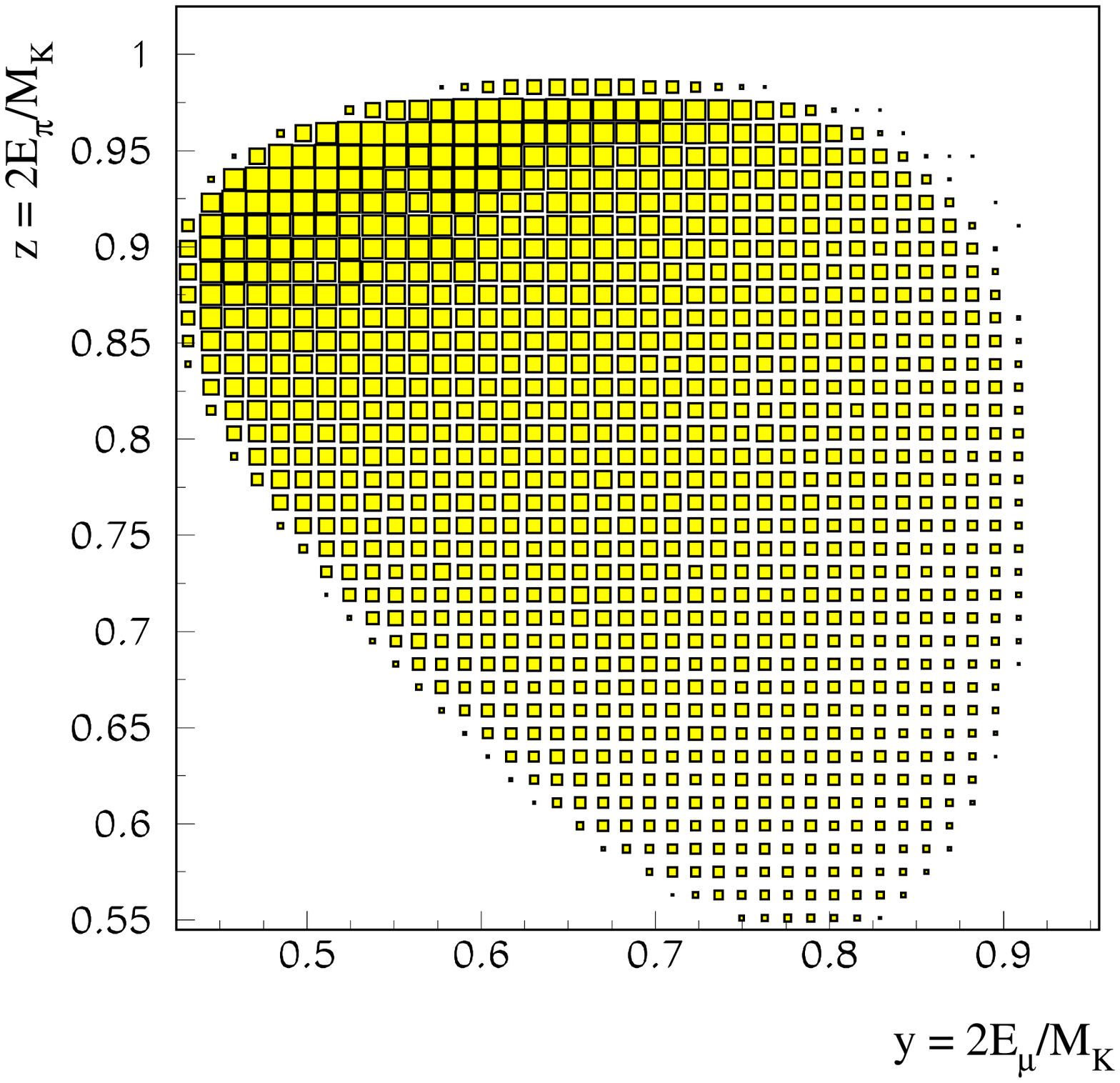,width=9cm}
\end{center}
\begin{center}
Figure 6: Dalitz plot for the selected $K \rightarrow \mu \nu \pi^{0}$ events 
after the 2C fit.
\end{center}

The procedure of the extraction of the form-factor parameters
starts with the subdivision of the 
  Dalitz plot region 
$ y= 0.425 \div 0.955;\; z=0.545 \div 1.025$  into $40\times 40$ bins.

The signal MC was generated with the constant matrix element and we have to 
calculate the amplitude-induced weights during the fit procedure. One can
observe that 
 the Dalitz-plot density function $\rho (y,z)$ of (2) can be presented in the 
factorisable form, i.e  
\begin{equation}
\rho (y,z)=  \sum_{\alpha=1,18}{F_{\alpha}(\lambda _{+},\lambda_{+}^{'}
,\lambda _{0}, \lambda_{0}^{'}, f_{S}, f_{T}) \cdot K_{\alpha}(y,z)}, 
\end{equation}
where $F_{\alpha}$ are simple  bilinear functions of the form-factor parameters
 and $K_{\alpha}(y,z)$ are the
kinematic functions which are calculated from the MC-truth information.
For each $\alpha$, the sums of $K_{\alpha}(y,z)$ over events  are accumulated 
in the
Dalitz plot bins (i,j) to which the MC events fall after the reconstruction.
Finally, every bin in the Dalitz plot gets 18 weights $W_{\alpha}(i,j)$ and
the density function $r(i,j)$ which enters into the fitting procedure is 
constructed:
\begin{equation}
r(i,j)= \sum_{\alpha=1,18}{F_{\alpha}(\lambda _{+},\lambda_{+}^{'}
,\lambda _{0}, \lambda_{0}^{'}, f_{S}, f_{T}) \cdot W_{\alpha}(i,j)}
\end{equation}

 This method allows one to avoid the
systematic errors due to the ``migration'' of the events over the Dalitz plot
due to the finite experimental resolution and automatically takes into account
the efficiency of the reconstruction and selection procedures.

To take into account the finite number of MC events in the particular bin and
strong variation of the real data events over the Dalitz plot, we minimize a 
${-\cal L}$ function defined as \cite{Saclay}:
\begin{equation}
-{\cal L} = 2\sum_j n_j\ln\left[ \frac{n_j}{r_j}\left( 1-\frac{1}{m_j+1}\right)
 \right] + 2\sum_j (n_j+m_j+1)\ln\left(
\frac{1+\frac{r_j}{m_j}}{1+\frac{n_j}{m_j+1}}\right),
\end{equation}
where the sum runs over all populated bins, and $n_j$, $r_j$ and $m_j$ are the
number of data events, expected events and generated Monte Carlo events
respectively. For large $m_j$ Eq. (8) reduces to the more familiar expression
\begin{displaymath}
-{\cal L} = \sum_j [2(r_j-n_j) + 2 n_j\ln n_j/r_j]
\end{displaymath}

The minimization is performed by means of the ``MINUIT'' 
program \cite{Minuit}.  The errors are calculated by ``MINOS'' procedure of 
``MINUIT'' at the level $\Delta {\cal L} = 1$, corresponding to 68\%\ 
coverage probability for 1 parameter.

\section{Results}

A  fit of the $K_{\mu 3}$ data with 
$f_{S}=f_{T}=\lambda_{+}^{'}=\lambda_{0}^{'}=0$ gives the following result
for $\lambda_{+}$ and  $\lambda_{0}$: $\lambda_{+}=0.0277 \pm 0.0013$;
$\lambda_{0}=0.0183 \pm 0.001$. The $\lambda_{+}-\lambda_{0}$ correlation
parameter is found to be $d \lambda_{0}/d \lambda_{+}=-0.348$.
The total number of
bins is 1054 and $\chi^2/\mbox{ndf} = 1.008$. The quality of the fit is
illustrated in figures 7 and 8 where the projected variables $y=2E_{\mu}/m_K$
and  $z=2E_{\pi^0}/m_K$ are presented.

\begin{minipage}[t]{8.cm}
\epsfig{file=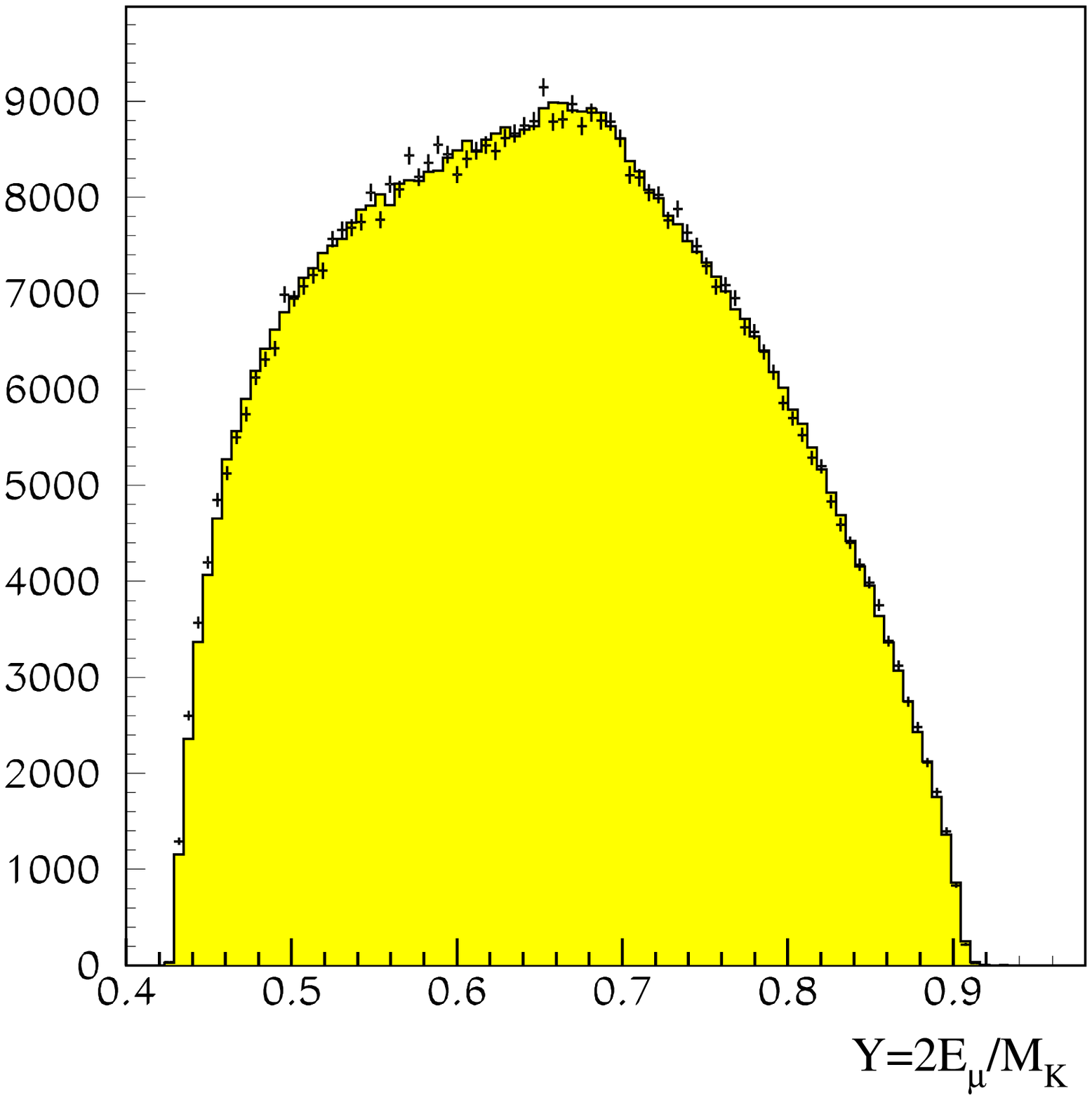,width=8cm}

\begin{center}
Figure 7: Y distribution. \\ The points with errors are the real data \\ and 
the shaded area -- signal MC.

\end{center}
\end{minipage} \ \hfill \ 
\begin{minipage}[t]{8.cm}
\epsfig{file=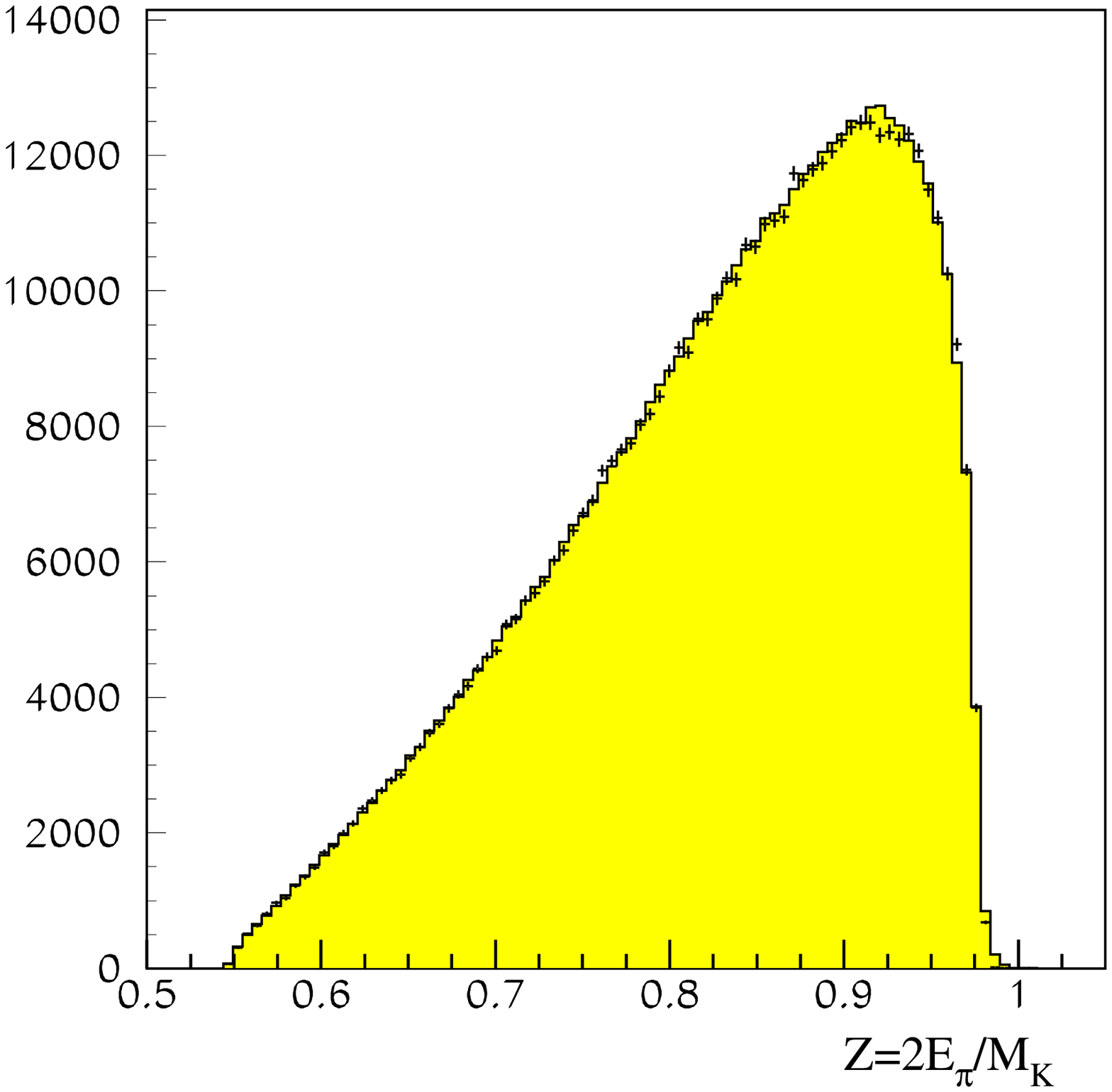,width=8cm}

\begin{center}
Figure 8: Z distribution. \\ The points with errors are the real data \\ and 
the shaded area -- signal MC.
\end{center}
\end{minipage}

~

~

 The value $\lambda^{\mu}_{+}=0.0277 \pm 0.0013$ is in a good agreement with 
that extracted from the analysis of our $K_{e 3}$ data \cite{papere}:
$\lambda^{e}_{+}=0.0286 \pm 0.00054(stat) \pm 0.0006(syst)$
(the statistical error $\pm 0.0008$ presented in \cite{papere} was 
obtained with $\Delta {\cal L} = 2.3$, corresponding to 68\%\ 
coverage probability for 2-parameter fit), 
 i.e, our data
do not contradict $\mu- e$ universality. 

In addition to the fits described above, Table 1 represents
the fits  with possible nonlinear 
terms in $f_+$ and $f_0$ (Eq. 4) as well as the fits with tensor and
scalar contributions (Eq. 1).

Every row of the Table 1 represents a particular fit where the parameters 
shown without errors are fixed.
The second row shows a fit where the nonlinearity is allowed in $f_{+}(t)$.
One can observe 
$\lambda_{+}-\lambda_{+}'$ correlation that results in the significant
$\lambda_{+}$ errors enhancement and visible shift of $\lambda_{+}$ and 
$\lambda_{0}$
parameters. The fitted value of $\lambda_{+}'$ is compatible with zero, while we
can not exclude some nonlinearity.
The third row represents a fit with the 
value of $\lambda_{+}'$ parameter extracted from the analysis of the 
data on pion scalar form-factors \cite{Bijnens}: 
$\lambda_{+}' = 3.2\cdot m_{\pi}^4 = 0.001063$. 
In a similar way 
  $\lambda_{0}'$ parameter is strongly correlated with 
$\lambda_{0}$ and is compatible with zero (row 4).

 We do not see any tensor contribution in our data (row 5).
The last row of the Table 1 represents a search for the scalar contribution.
As one can see from the Eq. (2), the $f_{S}$ term is 100\% 
anti-correlated with V-A contribution $(m_{\mu}/2m_{K})f_{-}$, where 
$f_{-}=f_{+}(0)(\lambda_{0}-\lambda_{+})\frac{m_{K}^{2}-m_{\pi}^{2}}{m_{\pi}^{2}}$,
i.e an independent estimate of this term is necessary. 
 A possible way consists in
  fixing $\lambda_{0}$ at the value calculated in the 
$\mbox{O}(p^4)$ ChPT:  
$\lambda_{0}^{\mbox{th}}= 0.017 \pm 0.004$ \cite{Leutwyler}.
The error $\pm 0.004$ in the theoretical prediction induces an additional error
of $\pm 0.0053$ in $f_S/f_{+}(0)$.

\renewcommand{\arraystretch}{1.4}

\begin{center}
\begin{tabular}{|cccc|}
\hline
  $\lambda_{+}$, $\lambda_{0}$ &  $\lambda_{+}^{'}$, $\lambda_0^{'}$ & 
   $f_T/f_{+}(0)$, $f_S/f_{+}(0)$ &  Fit prob. \\ \hline\hline
  $0.0277\pm 0.0013$ & 0. & 0. & 0.425 \\
  $0.0183\pm 0.0011$ & 0. & 0. &   \\ \hline
  $0.0215\pm 0.0060$ & $0.0010\pm 0.0010$ & 0. & 0.451 \\
  $0.0160\pm 0.0021$ &  0. & 0. &  \\ \hline
  $0.0216\pm 0.0013$ & 0.001063 & 0. & 0.451 \\
  $0.0163\pm 0.0011$ &  0. & 0. &  \\ \hline
  $0.0276\pm 0.0014$ & 0. & 0. & 0.421 \\
  $0.0170\pm 0.0059$ & $0.0002\pm 0.0008$ & 0. &  \\ \hline
  $0.0276\pm 0.0014$ & 0. & $-0.0007\pm 0.0071$ & 0.422 \\
  $0.0183\pm 0.0011$ & 0. & 0. &   \\ \hline
  $0.0277\pm 0.0013$ & 0. & 0. & 0.421 \\
  0.017 & 0. & $0.0017\pm 0.0014$ &   \\ \hline
  \hline
\end{tabular}
\end{center}

~

\begin{center}
Table 1. The $K_{\mu 3}$ fits.
\end{center}

Different sources of systematics are investigated. 
We allow variations of the muon selection cuts, angular 
cut and 2C-fit probability cut. 
The Dalitz plot binning, signal and
background MC variations are also applied.

The resulting systematic uncertainties are as follows:

\begin{itemize}
\item $\Delta\lambda_+ = 0.0009$ and $\Delta \lambda_0 = 0.0006$;
\item $\Delta f_{T}/f_{+}(0) = 0.002$ and  $\Delta f_{S}/f_{+}(0) = 0.0009$
\end{itemize}

\section{Summary and conclusions}
The $K^{-}_{\mu 3}$ decay has been studied using in-flight decays of 25 GeV 
$K^{-}$ detected by the ``ISTRA+'' magnetic spectrometer. 

 The $\lambda_{+}$ parameter of the vector form-factor 
 is measured to be: 
\begin{center} 
 $\lambda_{+}=0.0277 \pm 0.0013\; (stat) \pm 0.0009\; (syst)$
\end{center}
\noindent 

 The $\lambda_{0}$ parameter of the scalar form-factor 
 is defined: 
\begin{center} 
 $\lambda_{0}=0.0183 \pm 0.0011\; (stat) \pm 0.0006\; (syst)$
\end{center}
\noindent 

 The comparison of the  $\lambda_{+}$ parameter with that obtained from 
 our $K_{e 3}$ data shows $e-\mu$ universality.

It is, at present, the best measurement of these parameters. It is in a 
reasonable
agreement with $\mbox{O}(p^4)$ ChPT prediction as well 
as with recent $\lambda_{0}$ 
measurements from the $\Gamma(K_{\mu 3}) / \Gamma(K_{e 3})$ ratio \cite{Horie}. 

Possible quadratic contributions in the vector and scalar form-factors are
compatible with zero, further studies are necessary to perform a detailed
comparison of our data with $\mbox{O}(p^6)$ ChPT calculations \cite{Bijnens}.

The limits on 
 possible tensor and scalar couplings are derived from the combined fit: 
\begin{center} 
 $f_{T}/f_{+}(0)=-0.0007 \pm 0.0071\; (stat) \pm 0.002\; (syst) ; $ \\[3mm]
 $f_{S}/f_{+}(0)=0.0017\pm 0.0014\; (stat) \pm 0.0009\; (syst) \pm 0.0053\;
 (theor) $ 
\end{center}

\vspace*{0.5cm}

The work
is  supported by the RFBR grant N03-02-16330. \\

\end{document}